\documentclass[12pt]{article}
\usepackage{amsfonts}
\usepackage{mathrsfs}
\usepackage{amsmath,amssymb,array}
\usepackage{graphicx}

\numberwithin{equation}{section}
\def\p{\partial}

\begin{document}

\begin{titlepage}
\renewcommand{\thefootnote}{\fnsymbol{footnote}}

\begin{center}

\begin{flushright}\end{flushright}
\vspace{2.0cm}

\textbf{\Large{Morse Potential, Contour Integrals, \\[0.2cm]
               and Asian Options}}\vspace{2cm}

\textbf{Peng Zhang} \\[0.5cm]

E-mail: pzhang@bjut.edu.cn\\[0.5cm]

\emph{Institute of Theoretical Physics, College of Applied Sciences, \\
      Beijing University of Technology, Beijing 100124, P.R.China}

\end{center}\vspace{2cm}

\centerline{\textbf{Abstract}}\vspace{0.5cm} Completeness of the
eigenfunctions of a quantum mechanical system is crucial for the
probability interpretation. By using the method of contour integral
we give proper normalized eigenfunctions for both discrete and
continuum spectrum of the Morse potential, and explicitly prove the
completeness relation. As an application we use our spectral
decomposition formula to study the problem of the pricing of an
Asian option traded in the financial market.

\vspace{1.5cm}

\end{titlepage}
\setcounter{footnote}{0}

\section{Introduction}

Since its first introduction \cite{M29} to quantum physics in 1929,
the morse potential as a solvable model has been studied extensively
by many different approaches, see e.g. \cite{IH51}-\cite{C04}. Most
of these works focus only on the discrete spectrum. In this paper we
will discuss the completeness issue of its eigenfunctions for both
discrete and continuum spectrum. Completeness is a very important
property of a quantum mechanical system. It is crucial for the
probability interpretation. To establish the completeness of a set
of eigenfunctions, it is necessary to properly choose their
normalization constants, which will give the correct integral
measure in the spectral decomposition of a square-integrable wave
function. The main purpose of the present paper is to do so for the
Morse potential, especially for its continuum eigenstates.

There is a very powerful method \cite{T62}, due to E. C. Titchmarsh,
to study the above-mentioned spectral expansion problem for a given
self-adjoint Sturm-Liouville problem. The advantage of this method
is the fact that it uses only classical complex analysis, which is
rather familiar to physicists, and does not require the knowledge of
the abstract operator theory in functional analysis. By using Cauchy
contour integrals, the Titchmarsh method will automatically gives a
complete system of proper normalized eigenfunctions. The poles
inside the contour correspond to the discrete spectrum, while the
contributions from the cuts represent the continuum spectrum. It can
even deal with subtleties related with self-adjoint extensions.

In this paper we will apply the Titchmarsh method to establish the
completeness of the eigenfunctions of the Hamiltonian with Morse
potential. In section 2 we quickly review some main points of the
Titchmarsh expansion theory. In section 3 we apply this method to
the Morse potential. We give the proper normalization for both
discrete and continuum eigenfunctions, and prove the completeness
relation explicitly. In section 4 we discuss an interesting
application of our result to the pricing of an Asian option in the
financial market.

\section{Review of the Titchmarch expansion theory}

In this section we will review some main points of the Titchmarsh expansion theory \cite{T62}.
For a stationary Schr\"{o}dinger-type equation with a real potential function $V(x)$
\begin{eqnarray}
-\frac{\,d^2\psi}{\,dx^2}+V(x)\,\psi(x)=\lambda\,\psi(x)\,, \label{Sch_eq}
\end{eqnarray}
we want to get the eigenvalue $\lambda$ and the corresponding eigenfunction $\psi_\lambda(x)$
satisfying some proper boundary conditions. The discrete eigenvalues are interpreted as
energy levels of possible bound states, while the continuum part corresponds to scattering
states. For the consistency of the probability interpretation of quantum mechanics, it is
crucial that the set of all eigenfunctions forms a complete basis of the Hilbert space.
Without this property, probability is not conserved and unitarity is lost.
If $x\in(a,b)$ with finite $a,b$ and the potential function $V(x)$ is regular at both endpoints,
this problem is called regular. However if one of or both of two points tend to infinity, or the
potential becomes singular there, the problem is called singular. Since we want to apply this
method to the Morse potential, we assume that $x\in(-\infty,+\infty)$.

Define functions $\phi(x,\lambda)$ and $\theta(x,\lambda)$ to be solutions of (\ref{Sch_eq})
with boundary conditions
\begin{eqnarray}
\phi(0,\lambda)=0\,,&& \phi'(0,\lambda)=-1\,;\\
\theta(0,\lambda)=1\,,&&\, \theta'(0,\lambda)=0\,.
\end{eqnarray}
We introduce a finite cutoff $b>0$ instead of $+\infty$, and consider the following mapping
\begin{eqnarray}
\mu_+(z;\,b,\lambda):=-\,\frac{\,\,\theta(b,\lambda)\,z+\theta'(b,\lambda)\,\,}{\phi(b,\lambda)\,z+\phi'(b,\lambda)}\,\,.
\end{eqnarray}
This fractional linear transformation maps the real axis $\mathrm{Im}z=0$ to a circle. It is proved by H. Weyl that,
under the limit $b\rightarrow+\infty$, the image of real axis in the $z$-plane either contracts to a limit point
or converges to a limit circle. The Weyl-Titchmarsh $m$-function $m_+(\lambda)$ to value of the limit point or any
point on the limit circle. Then it follows that
\begin{eqnarray}
\psi_+(x,\lambda)=\,\theta(x,\lambda)+m_+(\lambda)\,\phi(x,\lambda)
\end{eqnarray}
is a solution of (\ref{Sch_eq}) and normalizable at $+\infty$. Similarly, there is another $m$-function $m_-(\lambda)$
for the left endpoint such that
\begin{eqnarray}
\psi_-(x,\lambda)=\,\theta(x,\lambda)+m_-(\lambda)\,\phi(x,\lambda)
\end{eqnarray}
is normalizable at $-\infty$. It can also be shown that, if in the
limit point case, these are the only normalizable solutions for each
endpoint. Define the resolvent kernel $G(x,x';\lambda)$ as
\begin{eqnarray}
G(x,x';\lambda)&=&\frac{\,\,\psi_+(x,\lambda)\,\psi_-(x,\lambda)\,\,}{m_+(\lambda)-m_-(\lambda)}\,, \qquad (x\geq x')\,\,; \nonumber\\
&=&\frac{\,\,\psi_-(x,\lambda)\,\psi_+(x,\lambda)\,\,}{m_+(\lambda)-m_-(\lambda)}\,, \qquad (x< x')\,\,.
\end{eqnarray}
It is the unique solution of
\begin{eqnarray}
-\frac{\,d^2G}{\,dx^2}+(V(x)-\lambda)\,G=\,\delta(x-x')\,,
\end{eqnarray}
which is symmetric between $x$ and $x'$ and square-integrable with respect to each variable.
For any wave function $f(x)$ to be expanded, let
\begin{eqnarray}
\Phi_f(x;\lambda)\,=\,\int_{-\infty}^{\infty}G(x,x';\lambda)\,f(x')\,dx'\,,
\end{eqnarray}
which is the solution of the inhomogeneous equation
\begin{eqnarray}
-\frac{\,d^2\Phi}{\,dx^2}+(V(x)-\lambda)\,\Phi=f(x)\,.
\end{eqnarray}
The function $\Phi_f(x;\lambda)$
can be analytically continued to complex $\lambda$ with possible poles and branch points. For the sake
of simplicity and further applications in the next section, we assume the function $\Phi_f(x;\lambda)$
is meromorphic in the complex $\lambda$-plane cut open along the positive real axis $\lambda\geq0$.
\begin{figure}[t]
\centering
\includegraphics[width=13cm]{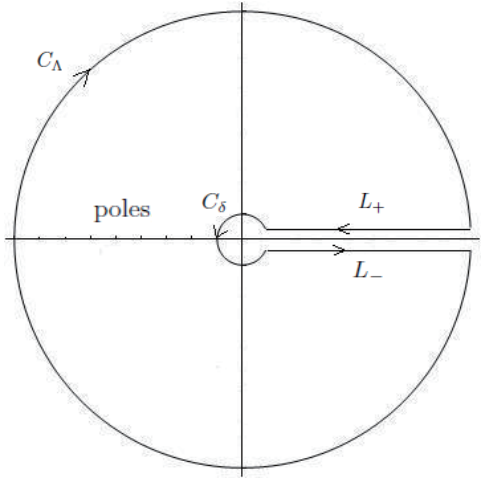}
\caption{\small{The integration contour $C:=\,C_\Lambda+L_++C_\delta+L_-$ which is used in (\ref{cont0}) and (\ref{cont1}).
We assume that $\lambda=0$ is a branch point, and on the negative real axis $\lambda<0$ there are some poles. We use the convention
about the argument: $\mathrm{arg}\lambda=0$ for $\lambda\in{L_-}$. }}\label{cont}
\end{figure}
By Cauchy's theorem we have
\begin{eqnarray}
\frac{1}{2\pi i}\oint_C \Phi_f(x;\lambda)\,d\lambda\,+\,\sum_{\mathrm{poles}}\,\mathrm{Res}\,\Phi_f \,=\,0 \,, \label{cont0}
\end{eqnarray}
where the contour $C$ is chosen as shown in Figure \ref{cont}.
For reasonable potential functions it can be proved that
\begin{eqnarray}
\frac{1}{2\pi i}\,\oint_{C_\Lambda} \Phi_f(x;\lambda)\,d\lambda\,\rightarrow\,f(x)\,, \qquad \Lambda\rightarrow\infty\,.
\end{eqnarray}
Therefore we obtain finally an equation as
\begin{eqnarray}
f(x)\,=\,-\sum_{\mathrm{poles}}\,\mathrm{Res}\,\Phi_f + \,\frac{1}{2\pi i}\int_0^\infty\hspace{-0.1cm}
\left(\Phi_f(x;\lambda\,e^{-2\pi i})-\Phi_f(x;\lambda)\,\right)d\lambda\,.
\end{eqnarray}
This is actually the desired expansion formula for wave function $f(x)$, and the completeness follows immediately.
From this quick review we can see the advantage of this method. It does not require any abstract operator theories,
and everything can be obtained just by the tools of classical complex analysis. So it is especially convenient for physicists.

\section{Spectral decomposition of the Morse potential}

Suppose $\kappa>0$, the Morse potential is \cite{M29}
\begin{eqnarray}
V_\mathrm{M}(x)=\,\kappa^2\left(\,e^{-2(x-x_0)}-2\,e^{-(x-x_0)}\,\right)\,. \label{Morse}
\end{eqnarray}
The corresponding Schr\"{o}dinger equation is
\begin{eqnarray}
-\frac{d^2\psi}{dx^2}+V_\mathrm{M}(x)\,\psi(x)=\lambda\,\psi(x)\,.\label{Sch}
\end{eqnarray}
Define
\begin{eqnarray}
u=2\kappa\,e^{-(x-x_0)}\,,\qquad
w(u)=\sqrt{u}\,\psi(x(u))\,,
\end{eqnarray}
we have
\begin{eqnarray}
\frac{d^2w}{du^2}+\left\{-\frac{1}{4}+\frac{\kappa}{u}+\frac{1/4-(-\lambda)\,}{u^2}\right\}w=0\,.
\end{eqnarray}
This is the Whittaker equation, so we have
\begin{eqnarray}
w(u)=\,c_1M_{\kappa,\,i\sqrt{\lambda}}\,(u)+\,c_2W_{\kappa,\,i\sqrt{\lambda}}\,(u)\,.
\end{eqnarray}
Therefore the solution of $\psi$ is
\begin{eqnarray}
\psi(x)&=&c_1\,\psi_1(x)+c_2\,\psi_2(x) \nonumber\\[0.2cm]
&\equiv&\frac{1}{\sqrt{2\kappa}}\,\,e^{(x-x_0)/2}\left[\,c_1M_{\kappa,\,i\sqrt{\lambda}}\,(2\kappa\,e^{-(x-x_0)})
+\,c_2W_{\kappa,\,i\sqrt{\lambda}}\,(2\kappa\,e^{-(x-x_0)})\,\right]\,.
\end{eqnarray}
The Wronskian $W(\psi_1,\psi_2)$ is
\begin{eqnarray}
W(\psi_1,\psi_2)\equiv\,\psi_1(x)\,\psi_2'(x)-\psi_1'(x)\,\psi_2(x)
=\,\frac{\Gamma(1+2\,i\sqrt{\lambda}\,)}{\,\Gamma(\,\frac{1}{2}-\kappa+i\sqrt{\lambda}\,)}\,\,.
\end{eqnarray}
When $x\rightarrow-\infty$, i.e. $u\rightarrow+\infty$, we have
\begin{eqnarray}
\psi_1&=&\frac{1}{\sqrt{u}}\,M_{\kappa,\,i\sqrt{\lambda}}\,(u)\,\sim\left(\,\alpha_+e^{u/2}u^{-\kappa}+\alpha_-e^{-u/2}u^{\kappa}\,\right)\,,\\
\psi_2&=&\frac{1}{\sqrt{u}}\,W_{\kappa,\,i\sqrt{\lambda}}\,(u)\,\sim\,e^{-u/2}u^{\kappa}\,.
\end{eqnarray}
When $x\rightarrow+\infty$, i.e. $u\rightarrow0$, we have
\begin{eqnarray}
\psi_1&=&\frac{1}{\sqrt{u}}\,M_{\kappa,\,i\sqrt{\lambda}}\,(u)\,\sim\,u^{i\sqrt{\lambda}}\,,\\
\psi_2&=&\frac{1}{\sqrt{u}}\,W_{\kappa,\,i\sqrt{\lambda}}\,(u)\,\sim\left(\beta_+u^{i\sqrt{\lambda}}+\beta_-u^{-i\sqrt{\lambda}}\,\right)\,.
\end{eqnarray}
If $-2\pi<\arg\lambda<0$, then $\mathrm{Im}\sqrt{\lambda}<0$, so
the solution normalizable at $-\infty$ is $\psi_2(x)$, while
the solution normalizable at $+\infty$ is $\psi_1(x)$.
Therefore both singular points are limit points.
The resolvent kernel $G(x,x';\lambda)$ is the normalizable solution of
\begin{eqnarray}
-\frac{d^2G}{dx^2}+(V(x)-\lambda)\,G=\delta(x-x')\,,
\end{eqnarray}
which is continuous at $x=x'$, but
\begin{eqnarray}
\left.\frac{\p G}{\p x}\,\right|_{x=x'+\varepsilon}-\left.\frac{\p G}{\p x}\,\right|_{x=x'-\varepsilon}=\,\,-1\,,\qquad \varepsilon\rightarrow0^+\,.
\end{eqnarray}
It can be checked that the solution is\footnote{In this paper we choose $-2\pi<\arg{\lambda}<0$. If we use the usual convention $0<\arg{\lambda}<2\pi$,
then $i\sqrt{\lambda}$ should be replaced by $-i\sqrt{\lambda}$ everywhere. These two different choices will give us the same
normalized eigenfunctions as in (\ref{Nd}) and (\ref{Nc}).}
\begin{eqnarray}
G(x,x';\lambda)&=&G_+(x,x';\lambda)\,\equiv\,\,\frac{\,\Gamma(\,\frac{1}{2}-\kappa+i\sqrt{\lambda}\,)}{\Gamma(1+2\,i\sqrt{\lambda}\,)}\,
           \frac{M_{\kappa,\,i\sqrt{\lambda}}\,(u)}{\sqrt{u}}\,\frac{W_{\kappa,\,i\sqrt{\lambda}}\,(u')}{\sqrt{u'}}\,,\quad x\geq x'\,;  \nonumber\\[0.1cm]
&=&G_-(x,x';\lambda)\,\equiv\,\,\frac{\,\Gamma(\,\frac{1}{2}-\kappa+i\sqrt{\lambda}\,)}{\Gamma(1+2\,i\sqrt{\lambda}\,)}\,\,
           \frac{W_{\kappa,\,i\sqrt{\lambda}}\,(u)}{\sqrt{u}}\,\frac{M_{\kappa,\,i\sqrt{\lambda}}\,(u')}{\sqrt{u'}}\,,\quad x<x'\,.\nonumber
\end{eqnarray}
This resolvent kernel $G(x,x';\lambda)$ has simple poles at
\begin{eqnarray}
\lambda_n=-\left(\kappa-n-1/2\right)^{\,2}\,,\qquad n=0,1,\cdots,[\kappa-1/2\,]\,.
\end{eqnarray}
which come form the factor $\Gamma(1/2-\kappa+i\sqrt{\lambda}\,)$ since\footnote{Note that the other factor
$M_{\kappa,\,i\sqrt{\lambda}}(u)/\,\Gamma(1+2\,i\sqrt{\lambda}\,)$ has no poles in the $\lambda$-plane.}
\begin{eqnarray}
\Gamma(1/2-\kappa+i\sqrt{\lambda}\,)\,=\,\frac{(-1)^{n+1}}{n!}\,\frac{2\,i\sqrt{\lambda_n}}{\lambda-\lambda_n}+\,O(1)\,.
\end{eqnarray}
Note that the upper limit $\kappa-1/2$ of $n$ follows from the requirement $\mathrm{Im}\sqrt{\lambda}<0$,
which is a direct sequence of $-2\pi<\arg\lambda<0$.\footnote{This upper limit does not depend on the convention
about the argument range of $\lambda$. If we choose $0<\lambda<2\pi$, it still holds.}
At these poles we have
\begin{eqnarray}
M_{\kappa,\,i\sqrt{\lambda_n}}\,(u)&=&\frac{n!\,\Gamma(2\kappa-2n)}{\Gamma(2\kappa-n)}\,e^{-u/2}u^{\kappa-n}L_n^{2\kappa-2n-1}(u)\,,\\
W_{\kappa,\,i\sqrt{\lambda_n}}\,(u)&=&(-1)^n\,n!\,e^{-u/2}u^{\kappa-n}L_n^{2\kappa-2n-1}(u)\,,
\end{eqnarray}
where $L_n^{\alpha}(u)$ is the generalized Laguerre polynomial.
Therefore the residues of $G(x,x';\lambda)$ at these poles are
\begin{eqnarray}
-\,\frac{\,n!\,(2\kappa-2n-1)}{\,\Gamma(2\kappa-n)}\,\,e^{-u/2}u^{\kappa-n-1/2}L_n^{2\kappa-2n-1}(u)\,\,
e^{-u'/2}u'^{\,\kappa-n-1/2}L_n^{2\kappa-2n-1}(u')\,.
\end{eqnarray}
For any function $f(x)$, the solution $\Phi_f(x;\lambda)$ of
$\Phi''+(\lambda-V(x))\,\Phi=f(x)$
is
\begin{eqnarray}
\Phi_f(x;\lambda)&=&\int_{-\infty}^{\infty}G(x,x';\lambda)\,f(x')\,dx' \nonumber\\
&=&\int_{-\infty}^{\,x}G_+(x,x';\lambda)\,f(x')\,dx'+\int_{x}^{\infty}G_-(x,x';\lambda)\,f(x')\,dx'\,.
\end{eqnarray}
As reviewed in the previous section, to obtain the spectral decomposition of the function $f(x)$,
we need to consider $\oint_C \Phi_f(x;\lambda)\,d\lambda$ with the contour $C$ defined in Figure \ref{cont}.
The Cauchy theorem tells us
\begin{eqnarray}
\oint_C \Phi_f(x;\lambda)\,d\lambda\,+\,2\pi i\sum_{\mathrm{poles}}\,\mathrm{Res}\,\Phi_f \,=\,0 \,. \label{cont1}
\end{eqnarray}
The contributions from poles, as calculated above, are
\begin{eqnarray}
&&-\,2\pi i\sum_{n=0}^{[\kappa-\frac{1}{2}]}\frac{\,n!\,(2\kappa-2n-1)}{\,\Gamma(2\kappa-n)}\,\,
       e^{-u/2}u^{\kappa-n-1/2}L_n^{2\kappa-2n-1}(u) \nonumber\\
&&\hspace{3.8cm} \times\int_{-\infty}^{\infty}e^{-u'/2}u'^{\,\kappa-n-1/2}L_n^{2\kappa-2n-1}(u')\,f(x')\,dx'\,.
\end{eqnarray}
Next we consider the sum of two integrals along the positive axis
\begin{eqnarray}
I_{L^+}+\,I_{L^-}&=&\int_{L^+}\Phi_f(x;\lambda)\,d\lambda\,+\int_{L^-}\Phi_f(x;\lambda)\,d\lambda \nonumber\\[0.1cm]
&=&\int_{-\infty}^{\,x}dx'f(x')\,\int_0^{\infty}d\lambda\left(\,G_+(x,x';\lambda)-G_+(x,x';\lambda\,e^{-2\pi i})\,\right)  \nonumber\\[0.1cm]
& &+\,\int_x^{\infty}dx'f(x')\,\int_0^{\infty}d\lambda\left(\,G_-(x,x';\lambda)-G_-(x,x';\lambda\,e^{-2\pi i})\,\right)  \nonumber\\[0.1cm]
&=&\int_{-\infty}^{\,x}dx'f(x')\,\int_0^{\infty}d\lambda\,
   \left(\frac{\,\Gamma(\,\frac{1}{2}-\kappa+i\sqrt{\lambda}\,)}{\Gamma(1+2\,i\sqrt{\lambda}\,)}\,
   \frac{M_{\kappa,\,i\sqrt{\lambda}}(u)}{\sqrt{u}}\right. \nonumber\\[0.1cm]
& &\hspace{2cm}\left. -\,\frac{\,\Gamma(\,\frac{1}{2}-\kappa-i\sqrt{\lambda}\,)}{\Gamma(1-2\,i\sqrt{\lambda}\,)}\,
   \frac{M_{\kappa,\,-i\sqrt{\lambda}}(u)}{\sqrt{u}}\right)\,\frac{W_{\kappa,\,i\sqrt{\lambda}}(u')}{\sqrt{u'}}  \nonumber\\[0.1cm]
& &+\,\int_{x}^{\infty}dx'f(x')\,\int_0^{\infty}d\lambda\,
   \left(\frac{\,\Gamma(\,\frac{1}{2}-\kappa+i\sqrt{\lambda}\,)}{\Gamma(1+2\,i\sqrt{\lambda}\,)}\,
   \frac{M_{\kappa,\,i\sqrt{\lambda}}(u')}{\sqrt{u'}} \right.  \nonumber\\[0.1cm]
& &\hspace{2.2cm}\left. -\,\frac{\,\Gamma(\,\frac{1}{2}-\kappa-i\sqrt{\lambda}\,)}{\Gamma(1-2\,i\sqrt{\lambda}\,)}\,
   \frac{M_{\kappa,\,-i\sqrt{\lambda}}(u')}{\sqrt{u'}}\right)\frac{W_{\kappa,\,i\sqrt{\lambda}}(u)}{\sqrt{u}}  \nonumber\\[0.1cm]
&=&\frac{1}{i\pi}\int_0^{\infty}d\lambda\,\sinh(2\pi\sqrt{\lambda}\,)\,
   \left|\,\Gamma\hspace{-0.1cm}\left(\frac{1}{2}-\kappa+i\sqrt{\lambda}\right)\,\right|^{\,2}
   \frac{W_{\kappa,\,i\sqrt{\lambda}}(u)}{\sqrt{u}} \nonumber\\[0.1cm]
&&\hspace{3cm}\times\int_{-\infty}^{\infty}\frac{W_{\kappa,\,i\sqrt{\lambda}}(u')}{\sqrt{u'}}\,f(x')\,dx'\,.
\end{eqnarray}
Here we have used the relation $W_{\kappa,-\mu}(u)=W_{\kappa,\,\mu}(u)$ and
\begin{eqnarray}
W_{\kappa,\,\mu}(z)=\,\frac{\Gamma(-2\mu)}{\,\Gamma\left(\frac{1}{2}-\kappa-\mu\right)}\,\,M_{\kappa,\,\mu}(z)+\,
                  \frac{\Gamma(2\mu)}{\,\Gamma\left(\frac{1}{2}-\kappa+\mu\right)}\,\,M_{\kappa,-\mu}(z)\,.
\end{eqnarray}
By (2.15.2) of \cite{T62}, for square-integrable function $f(x)$, we have
\begin{eqnarray}
\Phi_f(x;\lambda)=\,-\,\frac{f(x)}{\lambda}+\,O\left(\frac{1}{\,|\lambda|^{\frac{3}{4}}|\,\mathrm{Im}\lambda|\,}\,\right)\,,\qquad |\lambda\,|\rightarrow\infty\,.
\end{eqnarray}
So we have
\begin{eqnarray}
\lim_{\Lambda\rightarrow\infty}\,\int_{C_\Lambda}\Phi_f(x;\lambda)\,d\lambda\,=\,2\pi i f(x)\,.
\end{eqnarray}
It is easy to check that
\begin{eqnarray}
\lim_{\delta\rightarrow0}\,\int_{C_\delta}\Phi_f(x;\lambda)\,d\lambda\,=0\,.
\end{eqnarray}
Therefore  from the Cauchy's theorem (\ref{cont1}) we have
\begin{eqnarray}
f(x)&=&\sum_{n=0}^{[\kappa-\frac{1}{2}]}\frac{\,n!\,(2\kappa-2n-1)}{\,\Gamma(2\kappa-n)}\,\,e^{-u/2}u^{\kappa-n-1/2}L_n^{2\kappa-2n-1}(u) \nonumber\\
&&\hspace{1.3cm}\times\,\int_{-\infty}^{\infty}e^{-u'/2}u'^{\,\kappa-n-1/2}L_n^{2\kappa-2n-1}(u')\,f(x')\,dx' \nonumber\\[0.1cm]
&&+\,\,\frac{1}{\,2\pi^2}\int_0^{\infty}d\lambda\,\sinh(2\pi\sqrt{\lambda}\,)\,
   \left|\,\Gamma\hspace{-0.1cm}\left(\frac{1}{2}-\kappa+i\sqrt{\lambda}\right)\,\right|^{\,2}
   u^{-1/2}\,W_{\kappa,\,i\sqrt{\lambda}}(u) \nonumber\\[0.1cm]
&&\hspace{1.5cm}\times\,\int_{-\infty}^{\infty}u'^{-1/2}\,W_{\kappa,\,i\sqrt{\lambda}}(u')\,f(x')\,dx'\,, \label{exps}
\end{eqnarray}
with $u=2\kappa\,e^{-(x-x_0)}$ and similarly for $u'$.
We define the normalized eigenfunctions as\footnote{These eigenfunctions are actually real functions.}
\begin{eqnarray}
\psi_n(x)&=&\sqrt{\frac{\,n!\,(2\kappa-2n-1)}{\,\Gamma(2\kappa-n)}}\,\,\,e^{-u/2}u^{\kappa-n-1/2}L_n^{2\kappa-2n-1}(u)\,,\label{Nd}\\
\psi_\lambda(x)&=&\frac{1}{\sqrt{2}\,\pi}\,\sinh^{1/2}(2\pi\sqrt{\lambda}\,)\,
   \left|\,\Gamma\hspace{-0.1cm}\left(\frac{1}{2}-\kappa+i\sqrt{\lambda}\right)\,\right|\,u^{-1/2}\,W_{\kappa,\,i\sqrt{\lambda}}(u)\,, \label{Nc}
\end{eqnarray}
with $n=0,1,\cdots,[\kappa-1/2\,]$ and $\lambda>0$.
Then the expansion formula (\ref{exps}) leads to the completeness relation
\begin{eqnarray}
\sum_{n=0}^{[\kappa-\frac{1}{2}]}\psi_n^*(x)\,\psi_n(x')\,+\,\int_0^\infty\psi_\lambda^*(x)\,\psi_\lambda(x')\,d\lambda\,\,=\,\,\delta(x-x')\,.\label{Comp}
\end{eqnarray}
This is the main result of the present paper, which depends
crucially on the correct normalization (\ref{Nd}) and (\ref{Nc}) of
the eigenfunctions. The normalization of continuum states seems not
be discussed in the literature so far. In this section we have
derived it by the method of contour integral.

\section{Application to the Asian option pricing}

In this section we will apply the spectral decomposition formula (\ref{exps}) or (\ref{Comp})
to the problem of Asian option pricing. Let $S(t)$ denote the price of a stock and
$A(t)=\,\frac{1}{t} \int_0^{\,t} S(t)\,dt$
the time average up to the time $t$. An Asian put option with the strike price $K$ is a contract
which gives you the right to sale this stock at the expiration time $t=T$ and obtain the payoff $K-A(T)$ if $K>A(T)$,
while continue to hold the stock if $K<A(T)$. Therefore in either case the payoff is $P(T)=(K-A(T))^+$.\footnote{We use the definition
$X^+\hspace{-0.1cm}\equiv X\hspace{-0.1cm}\cdot\theta(X)$, where $\theta(\cdot)$ is the Heaviside step function.}
Obviously you should pay some money when signing
this option contract at $t=0$. Our aim is to determine its fair price.

Usually it is assumed that the stock price $S(t)$ follows the law of geometric Brownian motion, i.e.
$S(t)=S_0\,e^{\sigma W(t)+(r-\sigma^2/2)t}$, where $W(t)$ is a Brownian motion with constant volatility $\sigma$,
and $r$ is the risk-free interest rate.
After some rescalings the value of an Asian put option can be written as \cite{GY93}
\begin{eqnarray}
e^{-rT}\left(\frac{4S_0}{\sigma^2T}\right)\langle\,(k-a(\tau))^+\rangle\,,
\end{eqnarray}
where we have introduced the exponential functional $a(t)$ of the Brownian motion $W(t)$ as
$a(t)=\int_0^{t} e^{2(W(t)+\nu t)}dt$, together with dimensionless parameters $\tau=\sigma^2T/4,\,\nu=2r/\sigma^2-1$ and $k=\tau K/S_0$.
The probability density function $f(a,t)$ of the stochastic process $a(t)$ satisfies the following diffusion-type equation \cite{DGY01, L04}
\begin{eqnarray}
\frac{\p\hspace{-0.06cm}{f}}{\p{t}}=
\,2a^2\frac{\p^2\hspace{-0.08cm}{f}}{\p{a}^2}+(2(\nu+1)\,a+1)\,\frac{\p\hspace{-0.05cm}{f}}{\p{a}}\,\,,\quad a>0\,.
\end{eqnarray}
The heat kernel $K(a,a';t)$ is the solution of the above equation
with the initial condition $K(a,a';0)=\delta(a-a')$.
Then the option value can be written as
\begin{eqnarray}
e^{-rT}\left(\frac{4S_0}{\sigma^2T}\right)\int_0^\infty (k-a)^+ K(a,0;\tau)\,da\,.  \label{PO}
\end{eqnarray}
Suppose $f_\lambda$ satisfies
\begin{eqnarray}
-\,2a^2\frac{d^2\hspace{-0.08cm}{f_\lambda}}{d{a}^2}-
(2(\nu+1)\,a+1)\,\frac{d\hspace{-0.05cm}{f_\lambda}}{d{a}}=2\lambda\,f_\lambda\,, \label{diff}
\end{eqnarray}
which can be rewritten in the form $-(pf')'=\lambda wf$ by defining
$p(a)=a^{\nu+1}e^{1/(2a)}$ and $w(a)=a^{\nu-1}e^{-1/(2a)}$.
The heat kernel $K(a,a';t)$ has the following spectral decomposition
\begin{eqnarray}
K(a,a';t)=\,w(a)\,\sum_{\lambda}\, e^{-2\lambda t}  f_\lambda^*(a)f_\lambda(a')\,.
\end{eqnarray}
Introduce the new variable $x$ and new function $\psi_\lambda(x)$ as suggested in \cite{L04}
\begin{eqnarray}
x=\,\log{a}\,,\quad  \psi_\lambda(x)=\,a^{\nu/2}\,e^{-1/(4a)}f_\lambda(a)\,,
\end{eqnarray}
then the equation (\ref{diff}) can be reduced to
\begin{eqnarray}
-\,\frac{d^2\psi_\lambda}{d{x}^2}+\left(\frac{1}{16}\,e^{-2x}-\frac{1-\nu}{4}\,e^{-x}+\frac{\nu^2}{4}\right)\psi_\lambda=\,
{\lambda}\,\psi_\lambda\,.
\end{eqnarray}
Actually, up to a additional constant $\nu^2/4$, this is the static Schr\"{o}dinger equation
with the Morse potential. Compared with it standard form (\ref{Morse}), we have
\begin{eqnarray}
\kappa=\frac{1-\nu}{2}\,,\quad e^{x_0}=\,\frac{1}{2(1-\nu)}\,.
\end{eqnarray}
Here we restrict to the case $\nu<1$.
By use of our results (\ref{Nd}) and (\ref{Nc}) in the previous section, we can obtain the complete system
of eigenfunctions of (\ref{diff}) as follows
\begin{eqnarray}
f_n(a)&=&\sqrt{\frac{\,n!\,(-\nu-2n)}{2^{-\nu-2n}\,\Gamma(1-\nu-n)}}\,\,a^n L_n^{-\nu-2n}\left(\frac{1}{2a}\right)\,,\\[0.1cm]
f_\lambda(a)&=&\frac{1}{\pi}\,\,\sinh^{1/2}\left(2\pi\sqrt{\lambda-\nu^2/4}\,\right)\,
   \left|\,\Gamma\hspace{-0.1cm}\left(\nu/2+i\sqrt{\lambda-\nu^2/4}\,\right)\,\right|\, \nonumber\\[0.1cm]
&&\hspace{1cm}  \times\, a^{(1-\nu)/2}\,e^{1/(4a)}\,W_{(1-\nu)/2,\,i\sqrt{\lambda-\nu^2/4}}\left(\frac{1}{2a}\right)\,.
\end{eqnarray}
The corresponding eigenvalues are $\lambda_n=n(-\nu-n)$ with $n=0,1,\cdots,[-\nu/2]$,
together with the continuous spectrum $\lambda>\nu^2/4$.
Therefore, by redefining $p=2\sqrt{\lambda-\nu^2/4}$, we have
\begin{eqnarray}
K(a,0;\tau)&=&\sum_{n=0}^{[-\frac{\nu}{2}]}\,\frac{(-1)^n 2(-\nu-2n)}{\Gamma(1-\nu-n)}\,\,e^{-2n(-\nu-n)\tau}
   (2a)^{\nu+n-1} e^{-1/(2a)} L_n^{-\nu-2n}\left(\frac{1}{2a}\right) \nonumber\\[0.1cm]
&&+\,\,\frac{1}{2\pi^2}\int_0^{\infty} e^{-(p^2+\nu^2)\tau/2}(2a)^{(\nu-1)/2}\,e^{-1/(4a)} \nonumber\\
&&\hspace{1cm} \times\, W_{(1-\nu)/2,\,ip/2}\left(\frac{1}{2a}\right)
   \left|\,\Gamma\hspace{-0.08cm}\left(\frac{\nu+ip}{2}\,\right)\,\right|^{\,2}\sinh(\pi p)\,p\,dp \,.
\end{eqnarray}
We have used the asymptotic relation $L_n^\alpha(z)\sim\frac{(-1)^n}{n!}\,z^n$ and
$W_{\kappa,\,\mu}(z)\sim e^{-z/2}z^\kappa$ as $z\rightarrow\infty$.
Therefore the value of an Asian put option is
\begin{eqnarray}
&&e^{-rT}\left(\frac{4S_0}{\sigma^2T}\right)\,\left\{\,\sum_{n=0}^{[-\nu/2]}\frac{(-1)^n(-\nu-2n)}{2n!\,\Gamma(1-\nu-n)}\,\,
  e^{-2n(-\nu-n)\tau}(2k)^{\nu+n+3}\,e^{-1/(2k)}L_n^{-\nu-2n}\left(\frac{1}{2k}\right) \right.\nonumber\\
&&\hspace{3cm}+\,\,\frac{1}{8\pi^2}\int_0^{\infty}\,e^{-(p^2+\nu^2)\tau/2}(2k)^{\nu+3}\,e^{-1/(4k)} \nonumber\\
&&\hspace{4cm}\left.\times\,W_{-\frac{\nu+3}{2},\frac{ip}{2}}\left(\frac{1}{2k}\right)
  \left|\,\Gamma\hspace{-0.08cm}\left(\frac{\nu+ip}{2}\,\right)\,\right|^{\,2}\sinh(\pi p)\,p\,dp \,\right\}\,.
\end{eqnarray}
Here we have used the formula (7.623.7) of \cite{GR07} to work out the integration over $a$ in (\ref{PO}).
This is completely equivalent to the result of \cite{L04} which is obtained by a different method.

\section{Conclusions}

In this paper we obtain the properly normalized eigenfunctions of
the Morse potential and explicitly prove the completeness relation
by use of the method of contour integrals. In this method the
discrete spectrum corresponds the pole of the resolvent kernel,
while the continuum spectrum comes from the contributions of the
branch cut. We also apply this spectral decomposition to the problem
of an Asian option pricing. We expect this method can be applied to
more subtle spectral problems, e.g. issues about self-adjoint
extensions, in quantum mechanics.

\end{document}